\documentclass[twocolumn,showpacs]{revtex4}
\usepackage{amsmath}
\usepackage{epsf}

\begin{document}

\title{Asymmetric Landau-Zener tunneling in a periodic potential}
\author{M.~Jona-Lasinio, O.~Morsch, M. Cristiani, N.~Malossi,
   J.H.~M\"uller, E.~Courtade, M.~Anderlini  and E.~Arimondo}
\affiliation{INFM, Dipartimento di Fisica E.Fermi, Universit\`{a} di
Pisa, Via Buonarroti 2, I-56127 Pisa,Italy}

\date{\today}

\begin{abstract}
Using a simple model for nonlinear Landau-Zener tunneling
between two energy bands of a Bose-Einstein condensate in a periodic
potential,
we find that the tunneling rates for the two
directions of tunneling are not the same. Tunneling from the
ground state to the excited state is enhanced by the
nonlinearity, whereas in the opposite direction it is
suppressed. These findings are confirmed by numerical simulations of
the condensate dynamics.
Measuring the tunneling rates for a condensate of
rubidium atoms in an optical lattice, we have found
experimental evidence for this asymmetry.
\end{abstract}

\maketitle

The phenomenon of Landau-Zener (LZ) tunneling \cite{landzen} is a
basic quantum mechanical process. It is based on the solution of
the Schr\"odinger equation for a two-level dynamics when a
parameter of the Hamiltonian system is time dependent. If at time
$t=-\infty$ the system is prepared in one adiabatic state of the
Hamiltonian, the time dependence of the Hamiltonian implies that
at time $t=+\infty$ there is a finite probability that the system
will occupy the other adiabatic state. As far as this tunneling
behaviour is concerned, complete symmetry exists between the
adiabatic states. Variations of the LZ model have been studied
\cite{mullen-ben-tzt,gefen-ben-zt,fish-mul-benj},  and an
observation of LZ dynamics in classical optical systems has been
reported \cite{bou-dekker-lzdyn}. More recently, LZ tunneling
within a periodic potential was studied for a nonlinear two-level
system in which the level energies depend on the occupation of the
levels~\cite{zobay,wuniu-lz}. It was discovered that a
nonlinearity  with a positive sign enhances the tunneling
probability between the ground band and the first excited band.
Moreover, Niu and coworkers discovered a nonzero LZ tunneling
probability even in the fully adiabatic limit when the
nonlinearity was larger than a critical value~\cite{wuniu-lz}.
Critical values for deformations of the energy level structures
were obtained in refs.~\cite{zobay,pethick,mueller}. In a
Bose-Einstein condensate inside a periodic potential such as an
optical lattice, the mean-field interaction between the atoms can
be comparable to other energy scales of the system and hence the
level-dependent energy shift can lead to an observable
modification of the tunneling behaviour.

In the present work we explore, theoretically and experimentally,
the  Landau-Zener tunneling between Bloch bands of a Bose-Einstein
condensate in an accelerated optical lattice. The optical lattice
depth controls the tunneling barrier, while the optical lattice
acceleration controls the time dependence of the Hamiltonian. We
show that the mean-field nonlinearity produces an  asymmetry  for
the tunneling probability between the adiabatic states of the
Hamiltonian. More precisely the tunneling probability from the
lower energy adiabatic state to the upper one is enhanced, while
the tunneling probability from the higher energy level to a lower
level is suppressed. Numerical integration of the one-dimensional
Gross-Pitaevskii equation and a simple two-state model demonstrate
this asymmetry. Moreover, our experimental data obtained with a
rubidium Bose-Einstein condensate confirms this prediction.

The qualitative explanation of this asymmetry in the tunneling
transition probabilities depends on the choice of the quantum
representation used for the description of the nonlinear system.
We have identified two different basic mechanisms, acting
simultaneously or independently. In the first mechanism the
nonlinear term of the Schr\"odinger equation acts as a
perturbation whose strength is proportional to the energy level
occupation. If the initial state of the condensate in the lattice
corresponds to a filled lower level of the state model, then the
lower level is shifted upward in energy while the upper level is
left unaffected. This reduces the energy gap between the lower and
upper level and enhances the tunneling. On the contrary, if all
atoms fill the upper level then the energy of the upper level is
increased while the lower level remains unaffected. This enhances
the energy gap and reduces the tunneling. The overall balance
leads to an asymmetry between the two tunneling processes. In a
different representation the nonlinearity term in the
Schr\"odinger equation produces an additional term to the optical
lattice depth. This additional term increases or decreases the
optical lattice depth depending on whether the tunneling process
proceeds from the lower level to the upper one, or vice versa.
Thus the tunneling due to a weaker or stronger potential barrier
is increased or decreased, as the case may be, depending on the
initial state.\\ \indent The motion of a Bose-Einstein condensate
in an accelerated 1D optical lattice is described by the
Gross-Pitaevskii equation
\begin{multline} \label{schrod}
i\hbar \frac{\partial \psi}{\partial t} =
\frac{1}{2M}\left(-i\hbar\frac{\partial}{\partial
x}-Ma_{L}t\right)^2\psi + \\ + \frac{V_0}{2}\cos(2k_lx)\psi +
\frac{4\pi\hbar^2a_s}{M}\left|\psi\right|^2\psi
\end{multline}
where $M$ is the atomic mass, $k_L=\pi/d$ is the optical lattice
wavenumber with $d$ the optical lattice step, and $V_0$ is the
strength of the periodic potential depth. The $s$-wave scattering
length $a_s$  determines the nonlinearity of the system. Equation
\ref{schrod} is written in the comoving frame of the lattice, so
the inertial force $Ma_L$ appears as a momentum modification. The
wavefunction $\psi$ is normalized to the total number of atoms in
the condensate and we define $n_0$ as the average uniform atomic
density. Defining the dimensionless quantities $E_{rec}=\hbar^2
k_L^2/2M$, $\tilde x=2k_Lx$, $\tilde t=8E_{rec}t/\hbar$, and
rewriting $\tilde{\psi} =\psi/\sqrt{n_0}$, $\tilde
v=V_0/16E_{rec}$, $\tilde \alpha=M a_L/16E_{rec}k_L$, $C=\pi a_s
n_0/k_L^2$, Eq. \eqref{schrod} is cast in the following
form~\cite{wuniu-lz}:
\begin{multline} \label{schrod-adim}
i\frac{\partial \psi}{\partial t} = \frac{1}{2}\left(
-i\frac{\partial}{\partial x}-\alpha t \right)^2 \psi +v
\cos(x)\psi + C \left|\psi \right|^2 \psi
\end{multline}
where we have replaced $\tilde x$ with $x$, etc. In the
neighborhood of the Brillouin zone edge we can approximate the
wave function by a superposition of two plane waves (the two level
model of ref.~\cite{wuniu-lz}), assuming that only the ground
state and the first excited state are populated. We then
substitute $\psi(x,t)=a(t)e^{iqx}+b(t)e^{i(q-1)x}$, with
$|a(t)|^2+|b(t)|^2=1$ in Eq. \eqref{schrod-adim}. Comparing the
coefficients of $e^{iqx}$ and $e^{i(q-1)x}$, linearizing the
kinetic terms and dropping the irrelevant constant energy
$1/8+C[1+(|a|^2+|b|^2)/2]$, Eq. \eqref{schrod-adim} assumes the
form
\begin{multline}
\label{two-state-eq}
i\frac{\partial}{\partial t}\,\begin{pmatrix} a \\ b \end{pmatrix} =
   \left[\frac{\alpha t}{2}\sigma_{3}+\frac{v}{2}\sigma_{1}\right]
   \begin{pmatrix} a \\ b \end{pmatrix}+
\frac{C}{2}(|b|^2-|a|^2)\sigma_{3} \begin{pmatrix} a \\ b \end{pmatrix}
\end{multline}
where  $\sigma_i\,\,i=1,2,3$ are the Pauli matrices~\cite{note1}.
The adiabatic energies of Eq.~\ref{two-state-eq} have a butterfly
structure at the band edge of the Brillouin zone for $C \ge
v$~\cite{wuniu-lz,pethick,mueller}, but in the present work we
always work in a regime where $C\ll v$, hence that structure plays
no role.

\begin{figure}[htbp]
\centering\begin{center}\mbox{\epsfxsize 3.0 in
\epsfbox{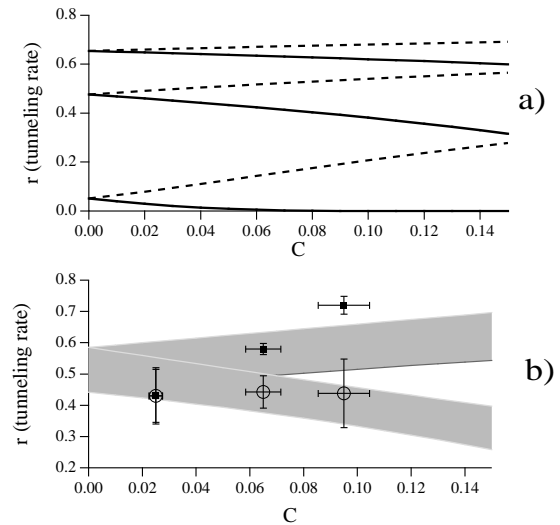}} \caption{LZ tunneling rate $r$ within the
two-level model as a function of the nonlinear parameter $C$ for
different accelerations. The rates coincide for $C=0$, whereas for
$C\neq0$ they differ significantly. Results are for $v=0.134$
corresponding to $V_{0}=2.2\,E_{rec}$, in (a) for
$\alpha=0.01,\,0.04,\,0.07$, corresponding to
$a_{L}=0.8,\,3.2,\,5.6\,\mathrm{m\,s}^{-2}$ from bottom to top and
in (b) for $\alpha=0.036$  corresponding to $a_{L}=2.9$ ms$^{-2}$.
In (b) experimental results are denoted by open symbols for the
excited band to ground state band tunneling and by filled symbols
for the ground state band to excited band tunneling. The
experimental points at $C=0.025$ have been re-scaled from
Fig.~\ref{Fig_LZlin} using the Landau-Zener formula to match the
acceleration to the one used to obtain the other experimental
points. The shaded area represents the confidence region for the
prediction of the two-level model taking into account the
uncertainty in our measurement of the lattice depth.
}\label{RateC}
\end{center}\end{figure}

In the linear regime $(C=0)$, evaluating the transition probability
in the adiabatic
approximation,
we find the linear LZ formula for the tunneling probability $r$
\begin{equation}
      r=e^{-\frac{\pi
v^2}{2\alpha}} \label{linearrate}
\end{equation}
expressing the occupation changes in terms of the rate $\alpha$ at
which the diagonal energies of the linear Hamiltonian change their
value, and of the off-diagonal interaction energy $v$. In the
nonlinear regime, as the nonlinear parameter $C$ grows, the lower
to upper tunneling probability grows as well until an adiabaticity
breakdown occurs at  $C=v$~\cite{wuniu-lz}. The upper to lower
tunneling probability, on the other hand, decreases with
increasing nonlinearity~\cite{note2}. We derived the tunneling
rate from the numerical integration of Eq.~\eqref{two-state-eq}.
In Fig.~\ref{RateC}(a) we plot the lower to upper tunneling rate
(initial $(a,b)$ =$(1,0)$) and the upper to lower tunneling rate
(initial $(a,b)$ =$(0,1)$) of the Bose-Einstein condensate as a
function of the nonlinear parameter $C$ for different
accelerations of the optical lattice. We see that for $C=0$ the
rate is the same for both tunneling directions whereas for $C\neq
0$ the two rates are different, and the smaller the acceleration
the larger the difference. This result is intuitive since for very
small accelerations the main contribution originates from the
nonlinear effect (the linear tunneling of Eq. (\ref{linearrate})
being small) while for large accelerations the main contribution
comes from the linear effect. We confirmed the presence of a
tunneling asymmetry by integrating directly Eq. \eqref{schrod}
(taking into account the full experimental protocol described
below), finding qualitative agreement with the prediction of the
two-state model. For small $C$ values, we have fitted the $C$
dependence of the tunneling rate of Fig.\ref{RateC} (a) through
the following expression~\cite{zobay,wuniu-lz}:
\begin{equation}
r(C)=e^{-\frac{\pi v^2}{2\alpha}(1\pm \beta\frac{C}{v})}
\end{equation}
with $\beta=0.75,0.17,0.14$ for the different acceleration values.

The nonlinear regime is more easily interpreted by
writing Eq. (\eqref{two-state-eq}) as
\begin{multline} \label{two-state1-eq}
i\frac{\partial}{\partial t}\,\begin{pmatrix} a \\ b \end{pmatrix} =
   \left[\frac{\alpha t}{2}\sigma_{3}+\frac{v}{2}\sigma_{1}\right]
   \begin{pmatrix} a \\ b \end{pmatrix}
- \frac{C}{2} \begin{pmatrix} |a|^2 & -b^*a \\
-a^*b & |b|^2 \end{pmatrix} \begin{pmatrix} a \\ b \end{pmatrix}
\end{multline}
Here the nonlinear diagonal terms represent a shift of the energy
levels, different for an occupation of the ground or upper level,
decreasing or increasing the energy gap depending on the initial
state.  The  off-diagonal terms of Eq.~\ref{two-state1-eq} produce
an equivalent contribution to the nonlinear tunneling process. The
off-diagonal terms modify the interaction term $v$ equivalent to a
Rabi frequency in the two-level model. The off-diagonal scalar
product between the two states $a^*b$, calculated applying the
adiabatic approximation technique of~\cite{crisp}, changes sign
depending on the initial state (lower or upper). Thus the linear
Rabi frequency is modified by the nonlinear off-diagonal term, and
the increase or decrease depends on the initial state. This
nonlinear modification of the interaction energy constitutes an
additional contribution to  the asymmetric tunneling rate.

Experimentally, we investigated the phenomenon of asymmetric
tunneling between the energy bands of Bose-Einstein condensates in
an optical lattice using a setup described in detail
in~\cite{morsch01,cristiani02}. Briefly, we create condensates of
$N\approx 10^4$ rubidium atoms in a time-orbiting potential (TOP)
trap. Once condensation has been achieved, the mean trapping
frequency $\overline{\nu}_{trap}$ of the magnetic trap is
adiabatically reduced to values between $15\,\mathrm{Hz}$ and
$50\,\mathrm{Hz}$. Thereafter, two laser beams with waists of
$1.8\,\mathrm{mm}$ and intersecting at an angle
$\theta=38\,\mathrm{deg}$ at the position of the condensate are
switched on with a linear ramp of duration
$\tau_{ramp}=10\,\mathrm{ms}$, thus ensuring adiabaticity of the
loading process. The beams are detuned to the red side of the
rubidium atomic resonance by $\approx 30\,\mathrm{GHz}$ and have a
variable frequency difference $\Delta \nu$ between them,
controllable through two acousto-optic modulators which are also
used to vary the intensity of the beams. In this way, a periodic
potential with lattice constant $d=1.18\,\mathrm{\mu m}$ and
lattice recoil energy $E_{rec}/h = 455\,\mathrm{Hz}$ is created,
which through the frequency difference $\Delta \nu$ can be made to
move at a constant velocity $v= d \Delta \nu$ or accelerated with
$a_{L} = d \frac{d\Delta \nu}{dt}$. For the current experiment,
lattice depths between $0.25\,E_{rec}$ and $2.5\,E_{rec}$ were
used.

\begin{figure}[htbp]
     \centering\begin{center}\mbox{\epsfxsize 2.65 in
\epsfbox{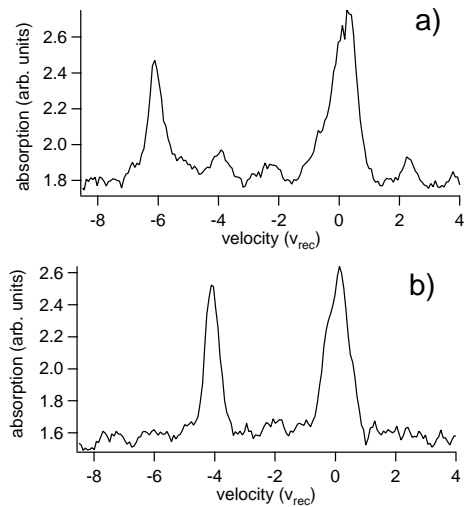}}
     \caption{Profiles of absorption images taken after
$22\,\mathrm{ms}$ of time-of-flight of condensates released after the
     acceleration procedure described in the text. The condensates were
prepared in the ground state band in (a) and in the first
     excited band in (b) within an optical lattice with depth
     $V_{0}=2.6\,E_{rec}$ and $a_{L}=2.9$ ms$^{-2}$.}\label{Fig_prof}
     \end{center}\end{figure}

Landau-Zener tunneling between the two lowest energy bands of a
condensate inside an optical lattice is investigated in the
following way. Initially, the condensate is loaded adiabatically
into one of the two bands. Subsequently, the lattice is
accelerated in such a way that the condensate crosses the edge of
the Brillouin zone once, resulting in a finite probability for
tunneling into the other band (higher-lying bands can be safely
neglected as their energy separation at the edge of the Brillouin
zone is much larger than the band gap). After the tunneling event,
the two bands have populations reflecting the Landau-Zener
tunneling rate (assuming that, initially, the condensate populated
one band exclusively). In order to experimentally determine the
number of atoms in the two bands, we then {\em increase} the
lattice depth (from $\approx 2\,E_{rec}$ to $\approx 4\,E_{rec}$)
and {\em decrease} the acceleration (from $\approx
3\,\mathrm{m\,s^{-2}}$ to $\approx 2\,\mathrm{m\,s^{-2}}$). In
this way, successive crossings of the band edge will result in a
much reduced Landau-Zener tunneling probability between the ground
state band and the first excited band (of order a few percent).
The fraction of the condensate that after the first tunneling
event populated the ground state band will, therefore, remain in
that band, whereas the population of the first excited band will
undergo tunneling to the second excited band with a large
probability (around $90$ percent) as the gap between these two
bands is smaller than the gap between the two lowest bands by a
factor $\approx 5$ for our parameters. Once the atoms have
tunneled into the second excited band, they essentially behave as
free particles since higher-lying band-gaps are
     smaller still, meaning that the fraction of the condensate that
     populated the first excited band after the first tunneling
     event will no longer be accelerated by the lattice. In summary,
     using this experimental sequence we selectively accelerate
     that part of the condensate further that populates the ground
     state band. In practice, in order to get a good
     separation between the two condensate parts after a
     time-of-flight, we accelerate the lattice to a final velocity
     of $4-6\,v_{rec}$~\cite{footnote_finvel} and absorptively image
the condensate after
$22\,\mathrm{ms}$ (see Fig.~\ref{Fig_prof}).

     In order to investigate tunneling from the ground state band to
     the first excited band, we adiabatically ramped up the lattice
     depth with the lattice at rest and then started the
     acceleration sequence. The tunneling from the first excited to the
ground-state band
     is investigated in a similar way, except that in this case we
     initially prepare the condensate in the first excited band by
     moving the lattice with a velocity of $1.5\,v_{rec}$ (through the
frequency difference $\Delta \nu$ between the acousto-optic
     modulators) when
     switching it on. In this way, in order to conserve energy and
     momentum the condensate must populate the first excited band at
     a quasi-momentum half-way between zero and the edge of the
     first Brillouin zone~\cite{nist}. Thereafter, the same acceleration
     sequence as described above is used in order first to induce
Landau-Zener tunneling and then to separate the
     fraction of the condensate that underwent tunneling to the lowest band
from the one
     that remained in the first excited band. For both tunneling
     directions, the tunneling rate is measured as
     \begin{equation}
     r=\frac{N_{tunnel}}{N_{tot}},
     \end{equation}
     where $N_{tot}$ is the total number of atoms measured from the
     absorption picture. For the tunneling from the first
     excited band to the ground band, $N_{tunnel}$ is the number of
     atoms accelerated by the lattice, i.e. those detected in the
     final velocity class $4\,v_{rec}$, whereas for the inverse
     tunneling direction, $N_{tunnel}$ is the number of atoms
     detected in the $v=0$ velocity class.
      \begin{figure}[htbp]
     \centering\begin{center}\mbox{\epsfxsize 2.65 in
\epsfbox{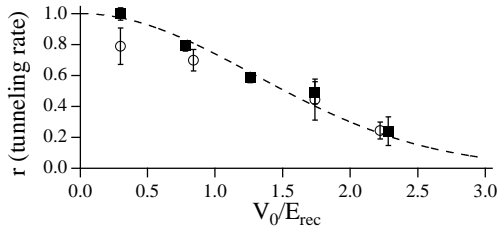}}
     \caption{Landau-Zener tunneling between the two lowest energy
bands of a condensate in an optical lattice as a function of
     the lattice depth (carrying a $\pm 10 \%$ systematic error) for $\alpha=0.025$ corresponding to $a_{L}=2.0$
     ms$^{-2}$. Tunneling rates
     from the ground state band to the first excited band (filled
symbols) and vice versa (open symbols) are virtually identical and
     agree with the linear prediction (dashed line) as the nonlinear
parameter $C\approx 0.025$ is small for the trap used in these
     measurements ($\overline{\nu}_{trap}=15\,\mathrm{Hz}$). The
deviation from theory of the filled
      symbol for small lattice depths is most likely due to a breakdown of the
adiabatic
      approximation of the linear Landau-Zener model. }\label{Fig_LZlin}
     \end{center}\end{figure}
\indent To illustrate our method for measuring the tunneling in
both directions, in Fig.~\ref{Fig_LZlin} we report the two
tunneling rates as a function of lattice depth for a condensate in
a weak magnetic trap and hence a small value of the interaction
parameter $C$~\cite{note3}. In this case, both tunneling rates are
essentially the same and agree well with the linear Landau-Zener
prediction. By contrast, when $C$ is increased, the two tunneling
rates begin to differ, as can be seen in Fig.~\ref{RateC}(b).
Qualitatively we find agreement with the theoretical predictions
of the non-linear Landau-Zener model, whereas quantitatively there
are significant deviations. We believe these to be partly due to
experimental imperfections. In particular, the sloshing (dipolar
oscillations) of the condensate inside the magnetic trap can lead
to the condensate not being prepared purely in one band due to
non-adiabatic mixing of the bands if the initial quasimomentum is
too close to a band-gap. Furthermore, a numerical simulation of
the experiment shows that for large values of $C$, for which the
magnetic trap frequency was large, the measured tunneling rates
are significantly modified by the presence of the trap. We have,
however, verified that when $C$ in the simulation is varied
without varying the trap frequency, the asymmetric tunneling
effect persists.\\ \indent In summary, we have numerically
simulated Landau-Zener tunneling between two energy bands in a
periodic potential and found that, in the presence of a nonlinear
interaction term, an asymmetry in the tunneling rates arises.
Experimentally, we have measured these tunneling rates for
different values of the interaction parameter and found
qualitative agreement with the simulations. We note here that the
phenomenon of asymmetric tunneling should be a rather general
feature of quantum systems exhibiting a nonlinearity. For
instance, calculating the energy shift due to a nonlinearity for
two adjacent levels of a harmonic oscillator, one finds that both
levels are shifted upwards in energy, the shift being proportional
to the population of the respective level. The energy difference
between the levels, therefore, decreases if only the lower state
is populated and increases if all the population is in the upper
level. Finally, we note that state-dependent mean-field shifts
have also been observed in measurements of the clock-shift in
ultra-cold and Bose-condensed atomic samples~\cite{harber02}.\\
\indent This work was supported by the MURST (COFIN2000
Initiative), the INFM (PRA 'Photonmatter'), and by  the EU through
Contract No. HPRN-CT-2000-00125.


\begin{thebibliography}{99}
\bibitem{note1} An equivalent equation is obtained using
wavefunctions $\exp(ikx)\cos(kx)$ and  $\exp(ikx)\sin(kx)$
describing the eigenstates at the band edge.
\bibitem{footnote_finvel} In order to keep the total
interaction time with the lattice constant, the final velocity of
the lattice when the condensate is prepared in the upper state is
$\approx 4\,v_{rec}$ since initially the lattice is moving at
$-1.5\,v_{rec}$.
\end{thebibliography}
\end{document}